\documentclass[twocolumn,showpacs,preprintnumbers,amsmath,amssymb]{revtex4-1}

\usepackage{graphicx}
\usepackage{natbib}
\usepackage{verbatim}
\usepackage{dcolumn}
\usepackage{textcomp}
\usepackage{amssymb}

\begin{document}

\begin{abstract}
We present the use of  direct bonded copper (DBC) for the straightforward fabrication of high power atom chips.  Atom chips using DBC have several benefits:  excellent copper/substrate adhesion, high purity, thick ($> 100~\mu$m) copper layers, high substrate thermal conductivity, high aspect ratio wires, the potential for rapid ($<8$~hr) fabrication, and three dimensional atom chip structures.  Two mask options for DBC atom chip fabrication are presented, as well as two methods for etching wire patterns into the copper layer.  The wire aspect ratio that optimizes the magnetic field gradient as a function of power dissipation is determined to be 0.84:1 (height:width).  The optimal wire thickness as a function of magnetic trapping height is also determined.  A test chip, able to support 100~A of current for 2~s without failing, is used to determine the thermal impedance of the DBC\@.  An assembly using two DBC atom chips to provide magnetic confinement is also shown.
\end{abstract}

\pacs{03.75.Be, 37.10.Gh, 41.20.Gz}

\title{Atom chips on direct bonded copper substrates}
\author{Matthew B. Squires}
\affiliation{Air Force Research Laboratory, Hanscom AFB, MA 01731, USA}
\email[]{AFRL.RVB.PA@hanscom.af.mil}

\author{James A. Stickney}
\affiliation{Air Force Research Laboratory, Hanscom AFB, MA 01731, USA}

\author{Evan J. Carlson}
\affiliation{Air Force Research Laboratory, Hanscom AFB, MA 01731, USA}

\author{Paul M. Baker}
\affiliation{Air Force Research Laboratory, Hanscom AFB, MA 01731, USA}

\author{Walter R. Buchwald}
\affiliation{Air Force Research Laboratory, Hanscom AFB, MA 01731, USA}

\author{Sandra Wentzell}
\affiliation{Air Force Research Laboratory, Hanscom AFB, MA 01731, USA}

\author{Steven M. Miller}
\affiliation{Air Force Research Laboratory, Hanscom AFB, MA 01731, USA}

\maketitle

Atom chips have been used in many cold atomic physics experiments including magneto-optical traps
\cite{art:MagneticSurfaceTrapsHansch}, magnetic traps \cite{art:MicroAtomTrapsLibbrecht,
  art:MagneticSurfaceTrapsHansch}, waveguides \cite{art:GuidingAtomsOnCurvesCornell, art:SpiralWaveguideSchmiedmayer},
transports \cite{art:ConveyorBeltHansch,art:CombinedAtomChipZimmermann}, Bose-Einstein condensates
\cite{art:ChipBECHansch}, coherent and incoherent beam splitters \cite{art:YBeamSplitterSchmiedmayer,
  art:RFdoubleWellInterferometrySchmiedmayer, art:MagneticBeamsplitterKetterle}, 
and have been integrated with optical elements \cite{art:BECInterferometerOnAChipCornell, art:ResonatorDetectionInChipVuletic,art:QEConChipStamperKurn}.  
Using atom chips for cold neutral atom physics has two main advantages.  First, by manipulating atoms $1-500~\mu\mbox{m}$ from
the chip wires, high magnetic field gradients can be generated at low power dissipation ($< 10$~W). Second, by using
well established lithographic fabrication techniques atom chip wires can be precisely placed to generate well
defined magnetic fields.  

For many applications, e.g. atom interferometry using light pulses, working distances of several hundred microns
are common.  As shown in Appendix~\ref{sec:OptimalAspectRatio}, power dissipation of a magnetic trap is minimized when the working
distance from the top surface of a rectangular wire is $z_0 = 0.7 h$ and when the aspect ratio is $h/w = \alpha = 0.8$, where $h$ is the thickness and
$w$ is the width of the wire.  Realization of thick $h \gtrsim 100 \mu\mbox{m}$  high quality metalizations with
standard lithographic techniques has proven to be technologically challenging, time consuming, and very expensive.  
For atom chip based magnetic traps that require working distances $z_0 \gtrsim 100~\mu\mbox{m}$, we present the use of direct bonded copper (DBC) substrates for atom chip fabrication. 

Direct bonded copper is the direct joining of a thin sheet of pure copper to a ceramic substrate
\cite{patent:BurgessDBC} and is commonly used in power electronics due to its high current handling and heat dissipation
properties.
Using DBC for atom chips addresses several atom chip fabrication requirements including power dissipation, ease of fabrication, thick metalizations, and excellent wire adhesion. 
The power dissipation of DBC atom chips is reduced for two reasons.  First, thick, pure copper layers are readily
available for DBC substrates with a resistivity equal to bulk copper.  Second, the ability to easily produce thick atom
chip wires that have a reduced net resistance.  Another benefit of using DBC for atom chips is the potential for rapid
in-house fabrication with minimal equipment.  We have produced moderate quality atom chips in less than eight hours, and
high quality wires in approximately one week with most of the time being taken by mask design and printing.  
By using DBC it is straightforward to fabricate various rectangular wire cross sections or aspect ratios because of the 
availability of a  thick copper layer.  Wires with aspect ratio $\alpha\gtrsim 2$ can be fabricated by wet etching,
and it may be possible to realize high aspect ratio wires $\alpha \sim 20$ using ablative laser micro-machining 
\cite{art:LaserMicroMachining}.

 To date, most atom chips have been fabricated using a Damscene process, where the wire height and  aspect ratio are determined by the
photoresist thickness and the  metal deposition (electroplating or evaporation) with typical wire heights on the order
of $10~\mu\mbox{m}$  \cite{art:WireTrapForNeutralsHansch,  art:AppsOfIntegratedMicrtrapsHansch,
  art:MicroElectroMagnetsPrentiss, art:WireImperfectionsAspect,art:AtomChipFabThermalPropertiesSchmiedmayer,
  art:CasimirPolderJohnsonNoiseBECVuletic, art:FastAtomChipBECHorikoshi}.  Thick lithographic wires have been fabricated
using thick photoresists \cite{art:GuidingAtomsOnCurvesCornell} or by depositing metal in channels formed in the
substrate (e.g. deep reactive ion etching in silicon) but the difficulty of fabrication has made thick wires less
common.  The thickness of a DBC layer is variable but is commonly as thick as 200 $\mu\mbox{m}$ for high current
applications.

Wire substrate adhesion is another key consideration when fabricating atom chips and this is especially true for thick wires because
residual stress from  metalization  increases the probability of delamination.  This problem is most often manifest
after the atom chip has been fabricated and is being prepared for use.  While this is a technological problem with a
known solution (i.e. careful surface preparation) it is a frequent issue that is solved by using DBC.  In our
experience, it takes considerable effort to remove the copper from the substrate which has made it possible to perform significant post processing of the atom chip.  For example, we have mechanically removed selected areas of the ceramic substrate after the copper has been etched so that the chip can be folded to yield a three dimensional ``origami chip''.

In this paper we will discuss basic fabrication methods, the resulting wire quality, and power dissipation.
  We will show DBC atom chips that are post-processed to make three dimensional chip structures that simplify electrical
  connections to the atom chip.

\section{Fabrication of Atom chips using Direct Bonded Copper}
To make DCB substrates, a copper sheet and a ceramic are bonded using the oxygen-copper eutectic ( 1.5\% oxygen at
$\sim1065\,^\circ \mbox{C}$ ) that is just below the melting point of pure copper ($\sim 1085 ^\circ\mbox{C}$).   During bonding, a thin
melt layer occurs at the oxide-copper interface that wets the ceramic surface and fills surface irregularities.  The
resulting bond is very strong because of the surface wetting between the copper and the oxide, and has  excellent
thermal conductivity because the intermediate liquid stage fills voids between the copper and the substrate.  During
bonding, the melt layer solidifies as diffusion of copper and oxygen at the interface moves the stoichiometric composition  of the
melt away from the eutectic transition.  To provide enough oxygen for the eutectic transition, a thin ceramic oxide layer
on the surface of the copper and ceramic is required.  The substrate can be an oxide (e.g. aluminum oxide
($\mbox{Al}_2\mbox{O}_3$) or  Beryllium oxide ($\mbox{BeO}$)) or
the surface can be oxidized prior to bonding (e.g. $\mbox{Al}_2\mbox{O}_3$ formation on aluminum nitride ($\mbox{AlN}$) or copper oxide on copper), however, the
surface oxidation step can reduce the bulk thermal conductivity due to the diffusion of oxygen along grain boundaries
\cite{book:PowerElectronicModulesDBC}. 

While any oxide can be used for DBC, most applications use electrically  insulating substrates that are thermally
conductive.  Beryllium oxide has the highest thermal conductivity ($260\, W/\mbox{m}^\circ\mbox{C}$) of the ceramics, but is not
commonly used because of beryllium toxicity.  Most DBC substrates are 
$\mbox{Al}_2\mbox{O}_3$ or $\mbox{AlN}$.    Aluminum oxide  is commonly used despite a moderate thermal conductivity
($\sim 25 \,\mbox{W/m}^\circ\mbox{C}$) because of its low cost, high substrate quality, and high bond strength.   Aluminum nitride is also used because
of its high thermal conductivity ($170\, W/\mbox{m}^\circ\mbox{C}$) and a coefficient of thermal expansion ($\sim4.7\, \mu\mbox{m/m}^\circ\mbox{C}$)
that is closely matched to silicon. 

Aluminum nitride has previously been used for atom chips with electroplated wires \cite{art:MagneticSurfaceTrapsHansch} where the
wires are in direct and conformal contact with the substrate,   however, electroplated metals typically have higher
resistivities than the bulk metal and, as a result, the maximum current density is reduced
\cite{art:EletroplatedCuFoo}.  Additionally, electroplated metals have a grained structure that results in variations of
the magnetic field close to the wire  \cite{art:WireImperfectionsAspect}.  In DBC, the bonded copper maintains the
electrical properties of the original copper sheet (i.e. high purity, high density, and low resistivity) for all
thicknesses of DBC\@.  The high purity of the DBC permits high total currents relative to electroplated wires of the same
thickness \cite{book:PowerElectronicModulesDBC}, or for the same current reduced power dissipation.   Pure and high
density atom chip wires have been deposited via evaporation with bulk resistivity and   smooth wire edges
\cite{art:AtomChipFabThermalPropertiesSchmiedmayer}.  The evaporation process is suitable for thin films ($<5~\mu\mbox{m}$) but
is complicated (e.g. refilling the evaporation source) and inefficient because only a small fraction of the evaporated
metal is deposited on the substrate while the remainder sticks to the walls of the vacuum system.

Excellent wire to substrate adhesion is a significant practical benefit of DBC atom chips.  Atom chips that use
electroplated or evaporated wires need an adhesion layer (e.g. $\mbox{Ti}$, $\mbox{TiW}$, $\mbox{Cr}$) to avoid delamination of the atom chip
wires.  The quality of the adhesion can vary significantly and depends on substrate cleanliness, deposition parameters,
and substrate compatibility.  Wire adhesion is especially important for thicker wires that may have residual stress from
the growth process.  The adhesion of DBC is excellent; it takes significant force to remove bulk copper
or thin wires from the substrate.  Because of the excellent adhesion, atom chips can be modified after patterning (see
Sec. \ref{sec:assembly}) while maintaining the integrity of the rest of the atom chip. 

High current hermetic vias is another option made possible by DBC and is a standard option for DBC substrates
\cite{art:DBCviasExel} and have been used as a part of a neutral atom trap \cite{art:DBC7LiBECKasevich}.  Commercially available vias
are incorporated into the substrate during the bonding process and require a custom manufacturing run.  It may also be
possible to form vias by drilling through the substrate and  soldering or electroplating in the hole to form a via.

Although there are many potential methods for building a DBC atom chip, the chips in this paper are fabricated by 
applying a mask to the surface of the DBC and then etching the copper away
from the exposed areas leaving wire traces on the AlN substrate.  For rapid prototyping (several hours), laser printer
toner is used for the mask and the copper is etched with hydrochloric acid plus dilute hydrogen peroxide.  For
$125 ~\mu\mbox{m}$ thick copper this method can reproducibly produce $\sim 175 ~\mu\mbox{m}$ wide wires with $\sim 500 ~\mu\mbox{m}$ center to center
spacing.  The minimum feature size is determined by the resolution of the laser printer and the thickness of the copper.  The
second fabrication method uses photoresist and a commercially printed mask with sub-micron resolution.  After the pattern transfer, using standard lithographic procedures,
 the wires are etched using either  hydrochloric/peroxide or reverse electroplating
\cite{diss:PowerElectronicsXingsheng}. For $125~\mu\mbox{m}$ thick copper this method can reproducibly produce wires widths of
$\sim 100~\mu\mbox{m}$ with $\sim 250~\mu\mbox{m}$ spacing.  With care wire widths of $\sim 50~\mu\mbox{m}$ with $\sim 200~\mu\mbox{m}$
spacing can be fabricated.  Specific processing details of each method follow below.

To promote mask adhesion, the DBC surface is first cleaned by scrubbing the surface with steel wool and then rinsing with
acetone and methanol.  For rapid mask production, laser printer toner provides a chemically resistant thermo-plastic that
adheres to a surface when it is melted.  The toner may be remelted for multiple transfers between different substrates.
A 600 dpi laser printer is used to print the chip design onto a commercially available PCB transfer film \footnote{An HP
  LaserJet printer was used for printing the mask.  Press-n-Peel and Pulsar toner transfer papers have been used with
  similar success.  Specific vendors are provided for information only and does not constitute endorsement.}.   The
transfer film will release the backing at the areas where the toner sticks to the DBC making it possible to remove the
original printing media.  The toner transfer procedure follows the manufacturer's instructions with special care given to
the temperature cycling because overheating will result in a distorted toner mask.  After etching, the mask is removed with acetone.  
The advantage of the toner transfer mask is rapid mask production with minimal equipment.  The feature size using toner
transfer is limited by the printer resolution, the spreading of the toner during transfer, and the DBC thickness. 

Improved mask quality and smaller wires are achieved using photolithography.  First, a
silicon nitride layer is applied to the DBC to improve photoresist adhesion.  A $20~\mu\mbox{m}$ photoresist layer (AZ4620 or SU8) is spin coated and then lithographically patterned. Compared to the toner transfer method, the
 photolithographically defined atom chip wires are significantly straighter and the achievable wire width is significantly narrower
($\sim50~\mu\mbox{m}$).  The sidewall definition and minimum line size is a strong function of the total copper thickness.
For the $125~\mu\mbox{m}$ thick DBC used in this paper, the narrowest line feature that could be reproducibly etched was
($\sim50~\mu\mbox{m}$).  Compared to a toner transfer produced atom chip a lithographically produced atom chip requires more time
(multiple days versus hours) and additional equipment, but is able to produce finer features, more consistent lines, and accurate
wire placement not possible using a toner mask.

Several test chips were etched using solutions of 37\% hydrochloric acid (HCl), hydrogen peroxide,
and water in a mixture of approximately 5:1:10 by volume.  The wire quality depended on the peroxide concentration, etching
time, and solution agitation.  The hydrogen peroxide concentration was varied from 2 to 0.2 by volume, with lower
concentrations yielding longer etching times (1-2 hrs) and the best wires.  Solution agitation reduces the etch times by
removing bubbles from the surface and provides fresh chemicals to the etching areas. At the end of the etching procedure
a pipette was used to provide local agitation to reduce wire footing and assure electrical isolation of the atom chip
wires.  For low concentrations of hydrogen peroxide (0.4 by volume) no agitation is necessary because of reduced surface
bubbles.  The acid/hydrogen peroxide etching method is isotropic and causes undercutting of the mask, resulting in a
trapezoidal wire shape and a wire width that is smaller than the original mask.  

\begin{figure}
\includegraphics[width=5.6cm]{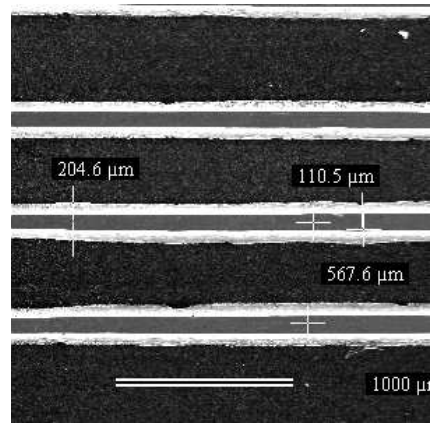}
\caption{A scanning electron micrograph of the top view of test chip A.\label{fig:TestATOP} }
\end{figure}
\begin{figure} 
\includegraphics[width=5.6cm]{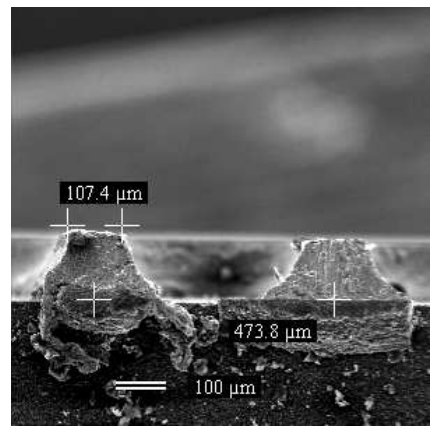}
\caption{A scanning electron micrograph of the cross sectional view of test chip A.\label{fig:TestASIDE} }
\end{figure}
The quality of various masking and etching methods was investigated by varying wire widths and spacings.  The resulting quality of the etching processes and masking methods were quantified using a scanning electron microscope.  Two test chips are presented below and are representative of the photolithographically defined process.  Figures \ref{fig:TestATOP} and \ref{fig:TestASIDE} are top and cross sectional views  of test chip
A.    This chip was fabricated using a photoresist mask (Hoechst AZ4620) that consisted of
equally spaced $200~\mu\mbox{m}$ wide traces on $125~\mu\mbox{m}$ thick DBC on AlN.  The copper was etched using a concentration of hydrogen peroxide of 1.2 parts by volume and the etching took about 15 min.   As shown in Fig. \ref{fig:TestATOP}, the traces produced by this method are straight, parallel, and the dimensions of each trace are nearly identical.  The footing of the wire is less uniform than the top of the wire because of varying etch rates at the end of the etching process.   Figure
\ref{fig:TestASIDE} is a cross sectional view of test chip A that shows a trapezoidal shape that is the result of the wet etching process used.   The width of the traces at the bottom is about $200~\mu \mbox{m}$ and the width at the top of the
trace is about $100~\mu\mbox{m}$.  The wet etch process used caused undercutting of approximately $50~\mu\mbox{m}$ from each side of the mask.  The
aspect ratio of these wires is about $\alpha \approx 1.25$.

\begin{figure}
\includegraphics[width=5.6cm]{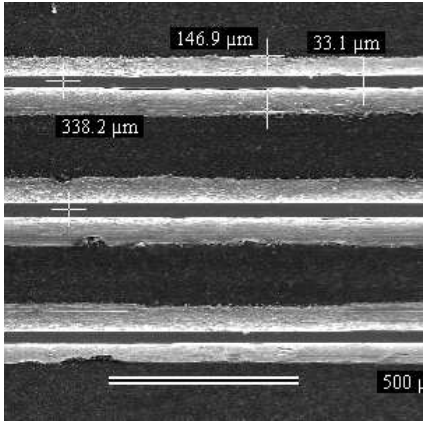}
\caption{A scanning electron micrograph of the top view of test chip B.\label{fig:TestBTOP} }
\end{figure}
\begin{figure}
\includegraphics[width=5.6cm]{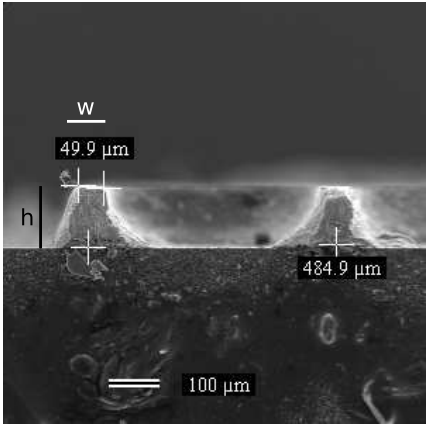}
\caption{A scanning electron micrograph of the cross sectional view of test chip B.\label{fig:TestBSIDE} }
\end{figure}
Figures \ref{fig:TestBTOP} and \ref{fig:TestBSIDE} are electron micrographs of test chip B.  This chip was made using the same substrate as test chip A, however, the hydrogen peroxide concentration was reduced to 0.6 parts by volume and the wire mask widths were $100~\mu\mbox{m}$.  The etching of this chip took about 30 min.  Similar to the $200~\mu\mbox{m}$ wide traces, Fig. \ref{fig:TestBTOP} shows the  $100~\mu\mbox{m}$ wide traces are also straight, parallel and reproducible. Figure \ref{fig:TestBSIDE} shows that for this chip,
the traces have straighter sidewalls and less undercutting than observed in test chip A.  This is attributed to the lower concentration of hydrogen peroxide.  The width of the top of
these wires was about $50~\mu\mbox{m}$ yielding an aspect ratio of $\alpha \approx 2.5$.

Test chips were also fabricated using reverse electroplating.  The DBC with applied photoresist mask was fixed parallel to a blank, similar sized
piece of copper and suspended in a copper sulfate solution with brighteners \footnote{Available from McMaster-Carr.  This
  is provided only as information and is not an endorsement.}.  A positive voltage was applied to the atom chip which
removed copper from the unmasked areas which was then deposited on the blank copper.  Low current densities (0.5~A/m$^2$)
produced the best results, but resulted in increased etching times.  A magnetic stirrer was used to provide fresh
solution to the surface of the DBC, however, for long etching times the photoresist mask debonded because of the
constant agitation of the electroplating solution.  Toward the end of the reverse electroplating, small ($\approx10\;\mu$m
diameter) copper islands remain because they were no longer electrically connected.  The copper islands were removed by a
rapid ($<5$~min) acid etch.   While reverse electroplating has the potential to provide higher aspect ratio wires and
close wire placement, the limits of the process are not determined at this time because of the mask integrity.

The maximum current handling and the thermal impedance of DBC atom chips was tested by fabricating a test chip with
three $1~\mbox{cm}$ long wires with different widths: $170~\mu\mbox{m}$,  $230~\mu\mbox{m}$, and  $340~\mu\mbox{m}$.  To
approximate the reduced thermal dissipation in vacuum, the chip was tested in air {\it without} a heat sink .   Only the
results of the thinnest wire are reported because the heating effects were found to be negligible for the wider wires.
The heating (measured using a thermocouple on the substrate $\sim$5~mm from the wire) and the voltage drop across the
wire were measured for currents of 10-100~A ($\sim0.4-5\times10^9$~A/m$^2$).  A 2~s current pulse of 100~A (110 W
average power) resulted in a 2.7 times increase in the wire resistance and $<15$~\textcelsius~ temperature rise.  For
short times ($<10$~ms) the thermal conductance of the wire/substrate junction was measured to be
$11\pm2\times10^6$~W/K-m$^2$, which is approximately 1.5-2 times the wire/substrate junction thermal conductance of
silicon with a 20~nm oxide layer \cite{art:AtomChipFabThermalPropertiesSchmiedmayer}.   With a suitable heat
sink the DBC atom chips should have equal or better performance  in vacuum.

\section{Assembly} \label{sec:assembly}

\begin{figure}
 \centering
 \includegraphics[width=0.4\textwidth]{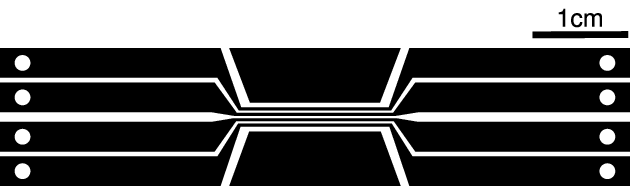}~ a.)
\vspace{.125 in}
 \includegraphics[width=0.4\textwidth]{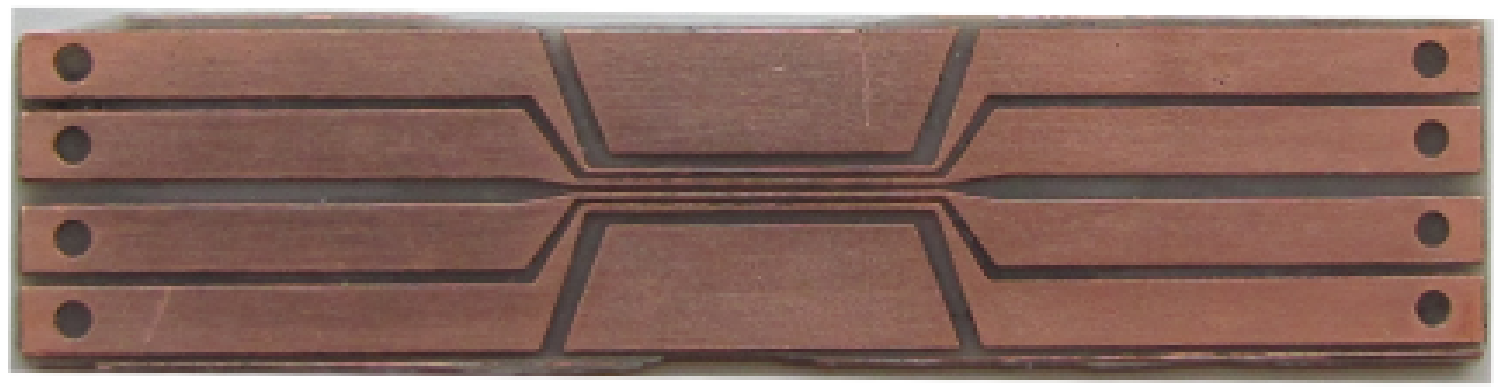}~ b.)
 \includegraphics[width=0.4\textwidth]{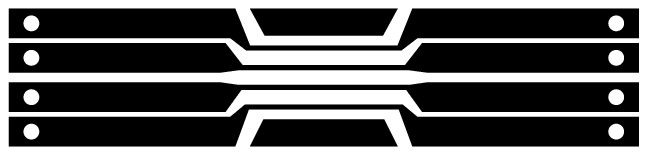}~ c.)
 \includegraphics[width=0.4\textwidth]{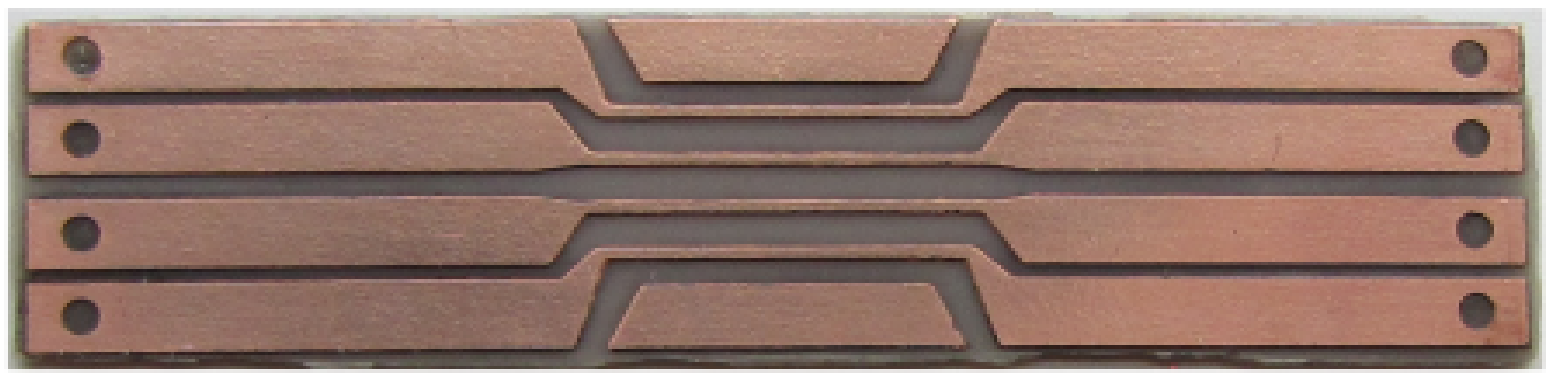}~ d.)
 \caption{Toner masks for etching the front a.) and back c.) chips and photographs of the etched DBC atom chips of front b.) and back d.) chips.}
 \label{fig:TonerMasksAndChips}
\end{figure}

We are currently developing a trapped atom interferometer gyroscope \cite{art:CollisonalDecoherenceStickney} that uses a cigar shaped harmonic trap.  Our plan is to realize this trap using a two layer DBC atom chip.  The chips were
fabricated using a toner transfer mask and a HCl/peroxide etch.    The chip closest to the
atoms (front chip) has four parallel wires that are $200~\mu\mbox{m}$ wide and spaced $500~\mu\mbox{m}$ center to
center.  Figure \ref{fig:TonerMasksAndChips}a shows the mask and Fig.~\ref{fig:TonerMasksAndChips}b is a photograph of
the final front chip.  The chip farthest from the atoms (back chip) has four parallel wires that are $400~\mu\mbox{m}$
wide and are separated by $1500~\mu\mbox{m}$.  The mask and chip for are shown in Figs.~\ref{fig:TonerMasksAndChips} c and d.
Areas of the chip that would otherwise be open are filled to reduce the amount of etchant used and to maintain a level top surface.  The chip wire leads are 2.5~mm wide to reduce power dissipation and provide strength when the substrate is removed in later processing.

After the atom chips were etched the substrates were modified for making electrical connections by drilling holes \footnote{We have use diamond drill bits from Dad's Rock Shop and diamond plated grinding bits from
  McMaster-Carr.  This is not an endorsement and is for information only.} at the end of the chip.  To improve optical access and to accommodate the space requirements, the edges of the chip were
folded back so the electrical connections could be made behind the mounting structure.  To avoid stressing and breaking the copper at the bend areas, the AlN substrate was removed
with a diamond grinding bit that is 2-3 times wider than the substrate thickness.  This is a straight forward and
reliable process because the diamond bit removes hard material significantly faster than soft materials (i.e. the AlN
can be removed without damaging the copper).  The copper traces were deformed by the slight pressure of the diamond
grinding bit but have no noticeable degradation in performance.  This  processing was made possible due to the excellent adhesion of the copper to the substrate and the copper thickness that is readily available using DBC.  This can also be accomplished by bonding multiple substrates with appropriate gaps, however this requires minimum lot runs and is not practical for one-off atom chip designs.

Perhaps the main advantage of foldable atom chips (dubbed ``origami chips'') is that they permit potentially bulky electrical connections to be made away from the surface of the atom chip.  This allows direct connections to be made to the chip while maintaining optical access over the full range of the chip.  High power electrical connections are available to the chip without bond wires, soldering, or vias.  The effect of the leads can be easily taken into account because the wires have a well defined path away from the chip.

\begin{figure}
 \centering
 \includegraphics[width=0.4\textwidth]{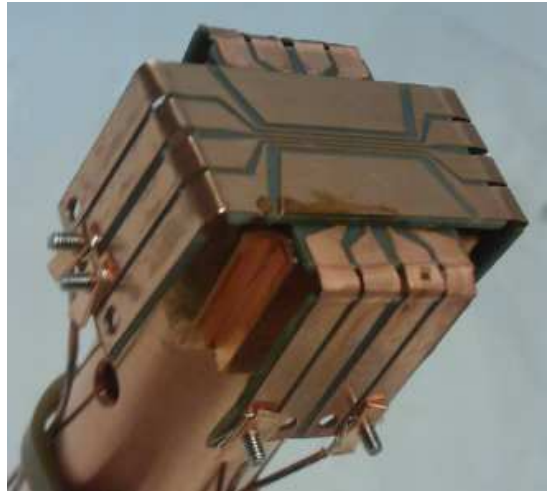}
\caption{Photograph of DBC atom chip assembly on copper mounting post.}
 \label{fig:ChipAssembly}
\end{figure}

Figure~\ref{fig:ChipAssembly} is a photograph of a DBC atom chip assembly constructed using two atom chips, shown
in Fig.~\ref{fig:TonerMasksAndChips}, that were processed, folded, stacked, and epoxied to a copper heat sink (using Epotek
353 ND).  Four of the wires were connected using vacuum compatible screws.  The assembly was placed in an UHV chamber and baked at $\sim130$~\textcelsius~ for approximately one week.  After
bakeout, the UHV chamber was isolated from the turbo pump and was  pumped solely by a 40 l/s ion pump to a final pressure of
$<2\times10^{-10}$ torr.  A $^{87}$Rb MOT was loaded directly below the DBC atom chip assembly and was then compressed
and cooled in a compressed MOT\@.  The atoms were magnetically trapped and transported to the DBC atom chip where they
were transferred to a chip quadrupole magnetic trap made by running current in the front chip wire plus an appropriate
bias field.  The ability to quantify the quality of the first chip was limited by the optical quality of the glass vacuum
chamber.  A new vacuum chamber has been constructed to address the optical quality issues.

\section{Conclusion}

Direct bonded copper substrates are an excellent candidate for atom chip fabrication because of thick high purity copper that is intimately bonded to a high thermal conductivity substrate.  Direct bonded copper atom chips can support very high currents (at minimum 100~A) for several seconds for high magnetic field gradient traps or for reduced power dissipation at lower currents.  Wires with the appropriate cross section will further optimize the magnetic field gradient for a given power dissipation.  The fabrication of DBC atom chips is straight forward and requires very little  equipment.  Direct bonded copper atom chips are very robust and can be bent to form 3D atom chip structures that simplifies the electrical connections away from the chips or for precision wire placement in multiple dimensions.

While this paper demonstrates the initial design and use of DBC for atom chips, there are multiple avenues for future improvements, such as, other copper patterning techniques (laser etching or dry etching \cite{art:CopperDryEtchLee})  which could produce higher density and/or higher aspect ratio wires.  Additionally, high precision and high current coils can be made for precise generation of centimeter size magnetic fields.  The DBC can also be patterned as a thermally conductive carrier for atom chips with finely patterned lithographic wires \cite{art:CombinedAtomChipZimmermann}.  The ability to form vias in the DBC also allows for greater flexibility in chip designs and for the potential to make high power hermetic feedthroughs.

\begin{acknowledgments}
This work was funded under the Air Force Office of Scientific Research under program/task 2301DS/03VS02COR.
\end{acknowledgments}

\appendix

\section{Optimal wire aspect ratio}
\label{sec:OptimalAspectRatio}

To determine the optimal aspect ratio that minimizes the power dissipated by a wire, consider the magnetic field
produced by a rectangular wire with height $h$ and width $w$ that carries current
in the $x$-direction.  In addition to the
field produced by the rectangular wire, there is a uniform bias field applied in the $y$-direction, such that there is a
minimum in the field along the line $y=0$, $z=z_0$.  Near this point, the magnitude of the magnetic field is
\begin{equation}
 \left|\mathbf{B}\right| \approx \sqrt{(B' y)^2 + (B' z)^2},
\end{equation}
and the magnetic field gradient is
\begin{equation}
 B'=\frac{J \mu_0}{\pi} \left[\arctan \left(\frac{2 (z_0+h)}{w}\right)-\arctan \left(\frac{2 z_0}{w}\right)\right],
\label{eqn:GradientCenteredWireHandW}
\end{equation}
where $J$ is the current density in the wire.  

In a typical experiment, the magnetic field gradient $B'$ is set at a particular value.  For example, in evaporative
cooling, the magnetic field must be large enough that the collision rate is high, but no so high that three body
collisions limit the lifetime of the trapped atoms.  Additionally, in chip experiments the distance from the center of
the trap to the chip $z_0$ is made as small as possible.  The minimum distance is typically chosen to avoid
fragmentation, the Casimir-Polder effect  \cite{art:CasimirPolderJohnsonNoiseBECVuletic}, thermal noise, or for optical access in atom
interferometry experiments.  Thus, the remaining task is to determine the height $h$ and width $w$ of a wire that
minimizes the power dissipation for a given magnetic field gradient $B'$ and working distance $z_0$.

The power dissipated by the current carrying wire is 
\begin{equation}
  \label{eq:Power}
  P = \rho l h w J^2,
\end{equation}
where $\rho$ is the resistivity and $l$ is the length of the wire.  Combining Eq.~(\ref{eqn:GradientCenteredWireHandW})
and Eq.~(\ref{eq:Power}) the dimensionless magnetic field gradient as a function of power dissipation can be written 
\begin{equation}
  \label{eq:fieldGrad_dimensionless}
  \beta = \zeta \alpha^{1/2} \left[
\arctan 2 \alpha (\zeta + 1) 
- \arctan 2 \alpha \zeta
 \right],
\end{equation}
where $\alpha = h/w$ is the aspect ratio of the wire, $\zeta = z_0 / h$ is the ratio of the trapping distance to the
wire thickness and 
\begin{equation}
  \label{eq:power}
  P = \rho l \left(
\frac{\pi z_0 B'}{\mu_0 \beta} 
 \right)^2,
\end{equation}
is the power dissipated by the wire.  Equation (\ref{eq:power}) shows that the power dissipation is minimized when $\beta$ is
maximized.  

\begin{figure}
  \includegraphics[width=8.6cm]{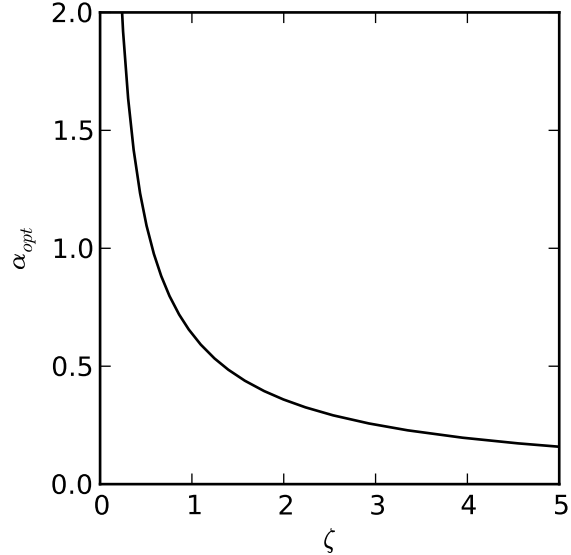}
  \caption{The optimal aspect ration $\alpha$ as a function of the ratio of trap height to wire thickness $\zeta$.}
  \label{fig:AlphaOPT_vs_Zeta}
\end{figure}

Analysis of Eq.~(\ref{eq:fieldGrad_dimensionless}) shows that for each value of $\zeta$, there is a value of $\alpha$
that maximizes the $\beta$.  Figure~\ref{fig:AlphaOPT_vs_Zeta} shows the optimal value of $\alpha$ as a function of
$\zeta$.  When $\zeta < 0.56$, the optimal aspect ratio is larger than one, which means that
when the trap is closer to the wire than the thickness of the wire.  This is because the magnetic field and corresponding gradient near a broad wire is constant.  In order to avoid the effects of a broad wire, the wire aspect ratio must be greater than approximately one.  On the other hand, when $\zeta > 0.56$, the optimal aspect ratio is smaller than one because the magnetic trap is far enough away to avoid the flattening of the magnetic field and the wider wire will result in a reduced wire resistance.

\begin{figure}
  \includegraphics[width=8.6cm]{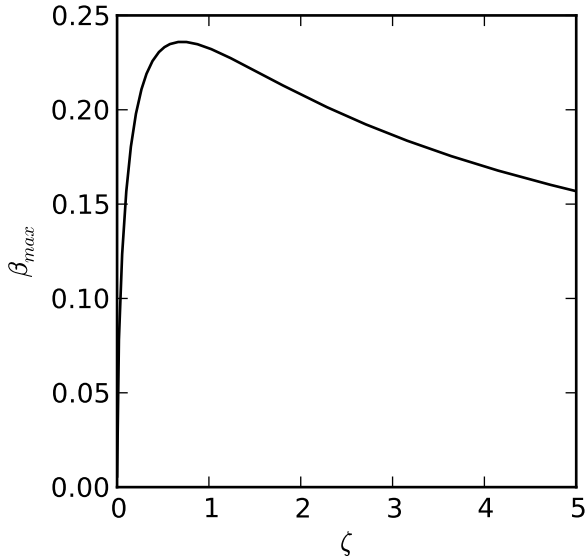}
  \caption{The maximum value of the dimensionless magnetic field gradient $\beta$ as a function of the ratio of trap
    height to wire thickness $\zeta$.}
  \label{fig:BetaMAX_vs_Zeta}
\end{figure}

Figure \ref{fig:BetaMAX_vs_Zeta} shows the maximum value of $\beta$ as a function of $\zeta$.  When $\zeta < 0.71$ the
maximum value of $\beta$ decreases rapidly reaching $\beta = 0$ when $\zeta = 0$.  Since the power dissipation is $P
\propto 1 /\beta^2$, the minimum power dissipation rapidly increases as the trap is brought close to the wire.  When
$\zeta > 0.71$, the maximum value of $\beta$ decreases, meaning that the minimum power dissipation increases as the trap
is moved far away from the wire. When $\zeta = 0.71$ and $\alpha = 0.84$ the power dissipation reaches a global minimum
with $\beta = 0.23$.

As an example, consider the first BEC on a chip \cite{art:ChipBECHansch} achieved using $\sim10\times50~\mu$m
gold wires electroplated on an AlN substrate.  The atoms were trapped $\sim 70~\mu$m from the chip surface.  Thus, $\alpha = 0.2$, $\zeta = 7$, and using Eq.~(\ref{eq:fieldGrad_dimensionless}) $\beta = 0.13$. For an optimal wire with $h = 100$ and $w = 120$ with the same field gradient and
trapping distance can be achieved while dissipating 0.3 times the power.

\bibliography{WireShapeOptimizedForGradient}

\end{document}